\documentclass[aps,prl,twocolumn,showpacs,groupedaddress,letterpaper,fleqn]{revtex4}  

\usepackage{graphicx}   
\usepackage{dcolumn}    
\usepackage{bm}         
\usepackage{multirow}   
\usepackage{subfigure}
\usepackage{amssymb}   
\usepackage{fleqn}

\newcommand{\etal}{\emph{et al.}}

\newcommand{\bsone}{\ensuremath{B_{s1}}}
\newcommand{\bstwo}{\ensuremath{B_{s2}^*}}
\newcommand{\bplus}{\ensuremath{B^+}}
\newcommand{\bst}{\ensuremath{B^{*+}}}
\newcommand{\jpsi}{\ensuremath{J/\psi}}
\newcommand{\bstwoM}{\ensuremath{5839.6}}
\newcommand{\bstwoMstat}{\ensuremath{1.1}}
\newcommand{\bstwoMsyst}{\ensuremath{0.7}}
\newcommand{\bstwoDM}{\ensuremath{66.7}}
\newcommand{\bstwoDMstat}{\ensuremath{1.1}}
\newcommand{\bstwoN}{\ensuremath{125}}
\newcommand{\bstwoNstat}{\ensuremath{25}}
\newcommand{\bstwoNsyst}{\ensuremath{9}}
\newcommand{\bstwoR}{\ensuremath{1.15}}
\newcommand{\bstwoRstat}{\ensuremath{0.23}}
\newcommand{\bstwoRsyst}{\ensuremath{0.13}}
\newcommand{\bplusN}{\ensuremath{20915}}
\newcommand{\bplusNstat}{\ensuremath{293}}
\newcommand{\bplusNsyst}{\ensuremath{200}}

\begin{document}


\hspace{5.2in} \mbox{Fermilab-Pub-07/585-E}

\title{Observation and properties of the orbitally excited {\bm $\bstwo$} meson}
%
\author{V.M.~Abazov$^{36}$}
\author{B.~Abbott$^{76}$}
\author{M.~Abolins$^{66}$}
\author{B.S.~Acharya$^{29}$}
\author{M.~Adams$^{52}$}
\author{T.~Adams$^{50}$}
\author{E.~Aguilo$^{6}$}
\author{S.H.~Ahn$^{31}$}
\author{M.~Ahsan$^{60}$}
\author{G.D.~Alexeev$^{36}$}
% ...

\affiliation{\vspace{0.1 in}(The D\O\ Collaboration)\vspace{0.1 in}}
% ...
\affiliation{$^{6}$University of Alberta, Edmonton, Alberta, Canada,
                Simon Fraser University, Burnaby, British Columbia, Canada,
                York University, Toronto, Ontario, Canada, and
                McGill University, Montreal, Quebec, Canada}
% ...
\affiliation{$^{29}$Tata Institute of Fundamental Research, Mumbai, India}
% ...
\affiliation{$^{31}$Korea Detector Laboratory, Korea University, Seoul, Korea}
% ...
\affiliation{$^{36}$Joint Institute for Nuclear Research, Dubna, Russia}
% ...
\affiliation{$^{50}$Florida State University, Tallahassee, Florida 32306, USA}
% ...
\affiliation{$^{52}$University of Illinois at Chicago,
                Chicago, Illinois 60607, USA}
% ...
\affiliation{$^{60}$Kansas State University, Manhattan, Kansas 66506, USA}
% ...
\affiliation{$^{66}$Michigan State University,
                East Lansing, Michigan 48824, USA}
% ...
\affiliation{$^{76}$University of Oklahoma, Norman, Oklahoma 73019, USA}
% ...
\date{November 2, 2007}

\begin{abstract}
We report the direct observation of the excited $L=1$ state $\bstwo$ in fully reconstructed decays to $\bplus K^{-}$. The mass of the $\bstwo$ meson is measured to be $\bstwoM \pm \bstwoMstat$~(stat.)~$\pm~\bstwoMsyst$~(syst.)~MeV/$c^2$, and its production rate relative to the $\bplus$ meson is measured to be $[\bstwoR \pm \bstwoRstat$~(stat.)~$\pm~\bstwoRsyst$~(syst.)$] \%$.
\end{abstract}

\pacs{12.15.Ff, 13.20.He,  14.40.Nd}

\maketitle

To date, the detailed spectroscopy of mesons containing a $b$ quark has not been fully established. Only the ground $J^P = 0^-$ states $\bplus$, $B^0$, $B^0_s$, $B^+_c$ and the excited $1^-$ state $B^*$ are established according to the PDG \cite{pdg}.
Previous studies of excited $(\bar b s)$ states have been carried out using inclusive final states with limited precision and ambiguous interpretation~\cite{opal}.
The properties of $(\bar b s)$ excited states, and comparison with the properties of the $(\bar b u)$ and $(\bar b d)$ systems, provide tests of various models of quark bound states and are important for their continuing development.

This Letter presents the observation of the process $\bstwo \to \bplus K^-$ with exclusively reconstructed $\bplus$ mesons, using a data sample corresponding to $1.3$ fb$^{-1}$ integrated luminosity collected with the D0 detector at the Fermilab Tevatron collider during 2002--2006. Charge conjugated states are implied throughout this Letter.

Quark models predict the existence of four P-wave ($L=1$) states in the $(\bar b s)$ system: two broad resonances ($B_{s0}^*$ and $B_{s1}^*)$ and two narrow resonances ($B_{s1}$ and $\bstwo$) \cite{hqs}. The broad resonances decay via S-wave processes and therefore are expected to have widths of a few hundred MeV/$c^2$. Such states are difficult to distinguish, in effective mass spectra, from the combinatorial background.
The narrow resonances decay via $D$-wave processes ($L=2$) and should have widths of approximately 1 MeV/$c^2$ ~\cite{falk}.  If the mass of the $B_{sJ}$ $(J=1,2)$ is large enough, the main decay channel should be $B_{sJ} \to B^{(*)} K$, since the $B_s \pi$ channel is forbidden by isospin conservation. A recent result by the CDF collaboration reports the observation of two narrow resonances consistent with the $\bsone$ and $\bstwo$ states~\cite{newcdf}.

The search for narrow $B_{sJ}$ mesons is performed by examining events with $B^{+(*)} K^-$ decays. This sample includes the following decays:
\begin{eqnarray}
\label{decay1}
B_{s1} & \to & \bst K^- ,~ \bst \to \bplus \gamma ; \\
\label{decay2}
\bstwo & \to & \bst K^- ,~ \bst \to \bplus \gamma ; \\
\label{decay3}
\bstwo & \to & \bplus K^- .
\end{eqnarray}
The direct decay $B_{s1} \to \bplus K^-$ is forbidden by conservation of parity and angular momentum. In decays (\ref{decay1}) and (\ref{decay2}), the photons from the $\bst$ decay have energy $E(\gamma) = (45.78 \pm 0.35)$ MeV \cite{pdg}. These photons are not reconstructed in this analysis, so that for such events the invariant mass of the reconstructed decay products is shifted down by $E(\gamma)$. 

The D0 detector is described in detail elsewhere~\cite{run2det}. The detector components most important for this analysis are the central
tracking and muon systems. The D0 central tracking system consists of a silicon microstrip tracker (SMT) and a central fiber tracker
(CFT), both located within a 2~T superconducting solenoidal magnet, with designs optimized for tracking and vertexing at pseudorapidities
$|\eta|<3$ and $|\eta|<2.5$, respectively (where $\eta$ = $-$ln[tan($\theta$/2)], and $\theta$ is the polar angle measured with respect to the proton beam direction). The muon system is located outside the calorimeters and has pseudorapidity coverage $|\eta|<2$.  It consists of a layer of tracking detectors and scintillation trigger counters in front of 1.8~T iron toroids, followed by two similar layers behind the toroids~\cite{run2muon}.

The data for this analysis were selected without any explicit trigger requirement, although most events satisfy inclusive single-muon triggers. The $\bplus$ mesons are reconstructed in the exclusive decay $\bplus \to J/\psi K^+$ with $\jpsi$ decaying to $\mu^+ \mu^-$. The selection procedure used is exactly as described in Ref.~\cite{bstst}. All $B$ mesons with mass $5.19 < M(\bplus) < 5.36$ GeV/$c^2$ are used, which yields a sample of $\bplusN \pm \bplusNstat$ (stat.)~$\pm 200$ (syst.) $\bplus$ candidates.

For each reconstructed $\bplus$ meson, an additional track with transverse momentum ($P_T$) above $0.6$ GeV/$c$ and charge opposite to that of the $\bplus$ meson is selected. This track is assigned the kaon mass.

For any track $i$, the significance $S_i$ is defined as 
$S_i = \sqrt{[\epsilon_T/\sigma(\epsilon_T)]^2 + [\epsilon_L/\sigma(\epsilon_L)]^2}$, 
where $\epsilon_T$ ($\epsilon_L$) is the projection of the track impact parameter on the plane perpendicular to the beam direction (along the beam direction), and $\sigma(\epsilon_T)$ [$\sigma(\epsilon_L)$] is its uncertainty.
Since the $B_{sJ}$ mesons decay at the production point, the additional track is required to originate from the primary vertex by applying the condition on its significance $S_K < \sqrt{6}$. The primary vertex is defined using the method described in Ref.~\cite{PV}.

\begin{figure}[t]
    \includegraphics[width=8.8cm]
{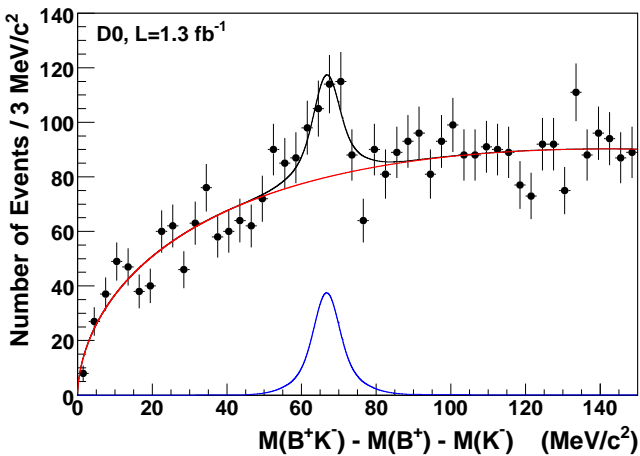}	
    \caption{Invariant mass difference $\Delta M = M(B^{+} K^{-}) - M(B^{+}) - M(K^-)$ for 
       exclusive $B$ decays. The line shows the fit described in the text, with signal and background contributions also plotted separately.}
    \label{bs2_fit}  
\end{figure}

For each combination satisfying the above criteria, the mass difference $\Delta M = M(B^{+} K^{-}) - M(B^{+}) - M(K^-)$ is computed. The resulting distribution of $\Delta M$ is shown in Fig.~\ref{bs2_fit}. 

Of the three decays (\ref{decay1}--\ref{decay3}) through which the $B_{sJ}$ states can reach the ground state $\bplus$, one or more may be kinematically suppressed if the excited state mass is smaller than the mass of the decay products. 
From inspection of Fig.~\ref{bs2_fit}, there is a single region of excess events above the background at $\Delta M = 67$ MeV/$c^2$, therefore the fit is based on the hypothesis that only one decay channel is observed. From kinematic considerations it follows that this is the highest energy transition, i.e. $\bstwo \to \bplus K^-$. Alternative hypotheses are discussed later.

Since the decay $\bstwo \to \bplus K^-$ occurs very close to the threshold $\Delta M = 0$ MeV/$c^2$, its width $\Gamma$ should be around 1~MeV/$c^2$~\cite{falk}. Because this is much less than the detector resolution, which is of order $6$~MeV/$c^2$, the fit is insensitive to values of $\Gamma$ below $6$~MeV/$c^2$, and $\Gamma$ is fixed at $1.0$ MeV/$c^2$. A systematic uncertainty is assigned to this choice of $\Gamma$ by fitting with a selection of small widths in the range $0$ to $2$ MeV/$c^2$.

Based on the above, the experimental distribution undergoes a binned maximum likelihood fit to the following function
\begin{eqnarray}
F(\Delta M) & = & F_{\text{sig}}(\Delta M) + F_{\text{bckg}}(\Delta M), \nonumber \\ 
F_{\text{sig}}(\Delta M) & = & N \cdot D(\Delta M, \Delta_0, \Gamma)
\label{fit}
\end{eqnarray}
In these equations, $\Delta_0$ is the central position of the resonance, i.e. $M(\bstwo) - M(\bplus) - M(K^-)$, $\Gamma$ is the $\bstwo$ width, and $N$ gives the total number of observed $\bstwo \to \bplus K^-$ decays. The background is parameterized by a modified power-law function:
\begin{eqnarray}
F_{\text{bckg}}(\Delta M) & = & c (\Delta M)^k + d (\Delta M),
\label{eq:bgfit}
\end{eqnarray}
where the parameters $c$, $d$ and $k$ participate in all fits. 

The function $D(\Delta M, \Delta_0, \Gamma)$ in Eq.~(\ref{fit}) is the convolution of a relativistic Breit-Wigner function with the experimental Gaussian resolution in $\Delta M$. The width of resonances in the Breit-Wigner function takes into account threshold effects using the standard expression~\cite{pdg,blwe} for $L=2$ decay.

The detector resolution function is determined from Monte Carlo simulation. All processes involving $B$ mesons are simulated using the {\sc EvtGen} generator \cite{evtgen} interfaced with {\sc pythia} \cite{pythia}, followed by full modeling of the detector response with {\sc geant} \cite{geant} and event reconstruction as in data. The difference between the reconstructed and generated values of $\Delta M$ is parameterized by a double-Gaussian function, with the width $\sigma_1$ ($\sigma_2$) of the narrow (wide) Gaussian set to 2.7 MeV/$c^2$ (6.2 MeV/$c^2$), and the normalisation of the narrow Gaussian set to $1.2$ times that of the wide Gaussian.
Studies of the $\bplus \to \jpsi K^+$ and $D^{*+} \to D^0 \pi^+$ decays show that simulation underestimates the mass resolution in data by $\approx$$10\%$. Therefore, the widths of the Gaussians which parameterise the $B_{sJ}$ resolution are increased by $10\%$ to match the data, and a $100\%$ systematic uncertainty is assigned to this correction.

Using a fitting range of $0 < \Delta M < 150$ MeV/$c^2$, covering $50$ bins, a binned maximum likelihood fit is performed. The following parameters of $\bstwo$ are obtained:
\begin{eqnarray}
\Delta_0   & = & M(\bstwo) - M(\bplus) - M(K^-) \nonumber \\
         & = & \bstwoDM \pm \bstwoDMstat~\mbox{(stat.) MeV/}c^2, \nonumber \\
      N  & = & \bstwoN \pm \bstwoNstat~\mbox{(stat.) events}. 
\label{fit_results}
\end{eqnarray}
Without the $\bstwo$ signal contribution, the log likelihood of the fit decreases by $13.4$, which implies that the signal is observed with a statistical significance of more than $5 \sigma$. 

To convert the $\Delta_0$ result into a mass measurement on $\bstwo$, the PDG values of the $\bplus$ ($5279.1 \pm 0.5$ MeV/$c^2$) and $K^-$ ($493.677 \pm 0.013$ MeV/$c^2$) masses are used as inputs \cite{pdg}. The uncertainties on these values are included in the systematic uncertainty on the $\bstwo$ mass. In addition, the mass is corrected by an amount $\epsilon_M$ to account for the D0 momentum scale uncertainty. This correction is in proportion to the difference between the mass of the $\bplus$ as measured by D0, and as listed by the PDG \cite{pdg}, leading to an upward shift in mass $\epsilon_M = +0.07$~MeV/$c^2$. A 100\% systematic uncertainty is assigned to this correction. Taking all factors into account, the mass $M(\bstwo)$ is measured to be:
\begin{eqnarray}
M(\bstwo) & = & \bstwoM \pm \bstwoMstat \pm \bstwoMsyst~\mbox{MeV/}c^2,
\label{abs_mass}
\end{eqnarray}
where the first uncertainty is statistical, the second systematic.

Taking the detected number of $\bplus$ ($\bplusN \pm \bplusNstat$) and $\bstwo$ ($\bstwoN \pm \bstwoNstat$) candidates, the production rate of $\bstwo$ relative to that of $\bplus$ is calculated as follows:
\begin{eqnarray}
R_J & = & \frac{Br(b \to \bstwo \to \bplus K^-)}{Br(b \to B^+)} 
     = \frac{N(\bstwo)}{N(\bplus) \cdot \varepsilon} \nonumber \\
    & = & (\bstwoR \pm \bstwoRstat \pm \bstwoRsyst) \%. 
\label{rate}
\end{eqnarray}
Here $\varepsilon$ is the relative detection efficiency of $\bstwo$ events compared to $\bplus$ events, i.e. it is the efficiency to select the additional kaon from the $\bstwo$ decay. The value of this parameter is determined from simulation to be $\varepsilon = 0.518 \pm 0.011$~(stat.), where the uncertainty results from the finite size of the simulation and is thus propagated into the measurement of $R_J$ as a systematic uncertainty. Emphasis is placed on agreement between the transverse momentum distributions in data and in simulation, and a systematic uncertainty is assigned to $\varepsilon$ to account for any difference.

Theoretical models predict that the $\bstwo$ meson, excluding phase-space factors, should decay with equal branching ratios into $\bst K$ and $\bplus K$. Decays into $\bst K$ will be observed as a resonance displaced to lower $\Delta M$ by the missing photon energy $45.78 \pm 0.35$ MeV~\cite{pdg}. An observation of this kind has already been made with the excited states of the $(\bar b d)$ quark system~\cite{bstst}.

Since the mass difference in the decay $\bstwo \to \bst K$ is very small, the rate should be strongly suppressed by a factor proportional to $(P^*/P)^5$, where $P^*$ ($P$) is the momentum  in the center-of-mass frame of the kaon in the decay $\bstwo \to \bst K$ ($\bplus K$)~\cite{falk}. Using the $\bstwo$ mass as measured here, a suppression factor of 0.074 is calculated; therefore no detectable $\bstwo \to \bst K$ signal is expected in the $\Delta M$ distribution with the current statistics.


To test for the presence of a $\bsone$ signal in the data, a two-peak hypothesis is used to fit the $\Delta M$ distribution. The $\bsone$ peak is assigned a physical width of $0$ MeV/$c^2$, and parameterized by a double-Gaussian function representing the experimental detector resolution. The resolution parameters are fixed from a separate simulation of $\bsone \to \bst K^-$ events. In this case, the widths $\sigma_{1,2}(\bsone)$ of the narrow and wide Gaussians are determined to be 1.1 and 2.2 MeV/$c^2$ respectively, and the normalisation of the narrow Gaussian is $3.6$ times that of the wide Gaussian. Again, the widths of the Gaussians are increased by $10\%$ to correct for underestimation in simulation. 

\begin{figure}[t]
    \includegraphics[width=8.8cm]
{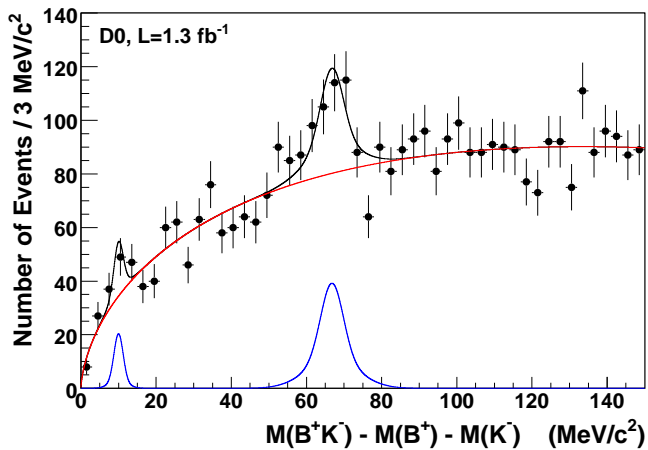}	
    \caption{Invariant mass difference $\Delta M = M(B^{+} K^{-}) - M(B^{+}) - M(K^-)$ for 
       exclusive $B$ decays. The line shows the fit with a two-peak hypothesis, as described in the text. Shown separately are contributions from signal and background.}
    \label{fig:bs1}  
\end{figure}

The resulting fit is shown in Fig.~\ref{fig:bs1}, giving the following parameters for the $\bsone$ signal:
\begin{eqnarray}
\Delta M(\bsone) & = & M(\bplus K^-) - M(\bplus) - M(K^-) \nonumber \\
                 & = & M(\bsone) - M(\bst) - M(K^-) \nonumber \\
                 & = & 11.5 \pm 1.4~\mbox{(stat.) MeV/}c^2, \nonumber \\
       N         & = & 25 \pm 10~\mbox{(stat.) events}. 
\label{fit:bs1_results}
\end{eqnarray}
Without the $\bsone$ signal contribution, the log likelihood of the fit decreases by $2.7$, which implies that this structure is observed with a statistical significance of less than $3 \sigma$. Hence with the current data, the existence of a $\bsone$ state can be neither confirmed nor excluded.

For the $\bstwo$ mass fit, the influences of different sources of systematic uncertainty are estimated by examining the changes in the fit parameters under a number of variations. The parameters describing the background are allowed to vary in the fit and their uncertainties are included in our results. The effect of binning is tested by varying the bin width and position. 
In addition, the fit is made without the $10\%$ mass resolution correction.
To check the effect of fixing the physical width $\Gamma$ of $\bstwo$ at 1.0 MeV/$c^2$, the fit is repeated with different widths in the range 0 -- 2 MeV/$c^2$.
The uncertainty in the absolute momentum scale, which results in a small shift of all measured masses, is assigned a 100\% systematic uncertainty. Finally, the uncertainties on the PDG masses of $\bplus$ and $K^-$~\cite{pdg} are propagated into the systematic uncertainty on the $\bstwo$ mass. The summary of all systematic uncertainties in the $\bstwo$ mass fit is given in Table~\ref{tab:sys1}. 
\begin{table}[th]
\caption{\label{tab:sys1}Systematic uncertainties of the $\bstwo$ parameters determined from the $\Delta M$
fit and from the conversion into the mass $M(\bstwo)$.
The rows show the various sources of systematic uncertainty as described in the text. The columns show the
resulting uncertainties for the two free signal parameters as described in Eq.~(\ref{fit}).}
\begin{ruledtabular}
\begin{tabular}{lcr}
Source                          & $\delta M(\bstwo)$ (MeV/$c^2$)&  $\delta N$        \\ \hline 
Bin widths/positions            & 0.3                           & 7                  \\
Value of $\Gamma$               & 0.3                           & 5                  \\
PDG mass uncertainties          & 0.5                           & 0                  \\
Momentum scale                  & 0.1                           & 0                  \\ 
Resolution uncertainty          & 0.1                           & 3                  \\ \hline
Total                           & $\bstwoMsyst$                 & $\bstwoNsyst$      \\ 
\end{tabular}
\end{ruledtabular}
\end{table}

\begin{table*}[th]
\caption{\label{tab:sysrate}Systematic uncertainties in the $\bstwo$ production rate measurement. 
The rows show the various sources of systematic uncertaintys as described in the text. 
The columns show the effect of these sources on the three parameters used in the $R_J$ measurement, and on the production rate itself.}
\begin{ruledtabular}
\begin{tabular}{lcccr}
source                              & $\delta[N(\bstwo)]$ & $\delta[N(\bplus)]$ & $\delta(\varepsilon)$ & $\delta(R_J) (\%)$ \\ \hline
$N(\bstwo)$ uncertainty             & $\bstwoNsyst$       & ---                 & ---                   & 0.08               \\
$N(\bplus)$ uncertainty             & ---                 & $\bplusNsyst$       & ---                   & 0.01               \\
Reweighting correction              & ---                 & ---                 & 0.002                 & 0.00               \\
Impact parameter resolution         & ---                 & ---                 & 0.022                 & 0.05               \\
Track reconstruction efficiency     & ---                 & ---                 & 0.036                 & 0.08               \\ 
Statistical effects from simulation & ---                 & ---                 & 0.011                 & 0.02               \\ \hline
Total                               & $\bstwoNsyst$       & $\bplusNsyst$       & 0.044                 & $\bstwoRsyst$   
\end{tabular}
\end{ruledtabular}
\end{table*}

The measurement of the relative production rate $R_J$ uses the kaon detection efficiency predicted in simulation, as well as the numbers of $\bstwo$ and $\bplus$ events.  
The systematic uncertainty on the number of $\bplus$ events, described in Ref.~\cite{bstst}, is $\pm \bplusNsyst$ events.
The systematic uncertainty on the number of $\bstwo$ events is $\pm \bstwoNsyst$ events (see Table~\ref{tab:sys1}). 

The uncertainty of the impact parameter resolution in the simulation is estimated to be $\approx$$10\%$ \cite{burdin2}.
It can influence the measurement of the selection efficiency of the kaon from the $\bstwo$ decay. To test for the effect of such an uncertainty, the efficiency is recalculated with the kaon impact parameter requirement varied by $\pm 10\%$. The resulting variation in efficiency is $\pm 0.022$.

The track reconstruction efficiency for particles with low transverse momentum is measured in Ref.~\cite{brat}, and good agreement between data and simulation is found. This comparison is valid within the uncertainties of branching fractions of different $B$ semileptonic decays, which is about 7\%. This uncertainty translates to an efficiency variation of $\pm 0.036$.
An additional systematic effect, associated with the difference in the momentum distributions of selected particles in data and in simulation, is taken into account. This yields an uncertainty in the efficiency of $\pm 0.002$.

Combining all these effects in quadrature, the total systematic uncertainty on the efficiency $\varepsilon$ is $0.042$. Both this and the statistical uncertainty $0.011$ on $\varepsilon$ must be propagated into the production rate measurement. The effects of contributions from the efficiency, and the number of detected $\bplus$ and $\bstwo$ candidates, are shown in Table~\ref{tab:sysrate}.

Different consistency checks of the observed signal are performed. Variation of the selection requirements on $P_T(K)$, $M(\bplus)$ and $S_K$, from $\bstwo \to \bplus K^-$ decays, does not produce any significant change in the results. Events with positively and negatively charged kaons are analyzed separately, and consistent results are obtained. 
A complementary sample of events containing a kaon not compatible with the primary vertex is selected, and no significant $\bstwo$ signal is observed. Events with wrong charge combinations ($B^{+}K^{+}$ and $B^{-}K^{-}$) also show a signal consistent with zero.

In conclusion, the $\bstwo$ state is observed in decays to $\bplus K^{-}$ with a statistical significance of more than $5 \sigma$. 
The measured mass is $\bstwoM \pm \bstwoMstat $(stat.)$ \pm \bstwoMsyst $(syst.) MeV/$c^2$. 
The $\bstwo$ relative production rate with respect to the $\bplus$ meson is $[\bstwoR \pm \bstwoRstat $(stat.)$ \pm \bstwoRsyst $(syst.)$] \%$.
Searching for a $\bsone$ signal gives inconclusive results with the currently available data set, which is expected to increase by a factor of five in the next few years.

%
We thank the staffs at Fermilab and collaborating institutions, 
and acknowledge support from the 
DOE and NSF (USA);
CEA and CNRS/IN2P3 (France);
FASI, Rosatom and RFBR (Russia);
CAPES, CNPq, FAPERJ, FAPESP and FUNDUNESP (Brazil);
DAE and DST (India);
Colciencias (Colombia);
CONACyT (Mexico);
KRF and KOSEF (Korea);
CONICET and UBACyT (Argentina);
FOM (The Netherlands);
Science and Technology Facilities Council (United Kingdom);
MSMT and GACR (Czech Republic);
CRC Program, CFI, NSERC and WestGrid Project (Canada);
BMBF and DFG (Germany);
SFI (Ireland);
The Swedish Research Council (Sweden);
CAS and CNSF (China);
Alexander von Humboldt Foundation;
and the Marie Curie Program.
%

\end{document}